\documentstyle[multicol,epsf,aps]{revtex}
\begin{document}
\title{Multiscaling in Infinite Dimensional Collision Processes} 
\author{E.~Ben-Naim$\dag$ and P.~L.~Krapivsky$\ddag$}
\address{$\dag$Theoretical Division and Center for Nonlinear Studies, 
Los Alamos National Laboratory, Los Alamos, NM 87545, USA}
\address{$\ddag$Center for Polymer Studies and Department of Physics,
Boston University, Boston, MA 02215, USA}
\maketitle
\begin{abstract} 
  We study relaxation properties of two-body collisions in infinite
  spatial dimension.  We show that this process exhibits multiscaling
  asymptotic behavior as the underlying distribution is characterized
  by an infinite set of nontrivial exponents. These nonequilibrium
  relaxation characteristics are found to be closely related to the
  steady state properties of the system.

\end{abstract}
\begin{multicols}{2} 
 
  Our understanding of the statistical mechanics of nonequilibrium
  systems remains incomplete, in sharp contrast with their equilibrium
  counterpart. The rich phenomenology associated with dynamics of far
  from equilibrium interacting particle systems exposes the lack of a
  unifying theoretical framework. Simple tractable microscopic models
  can therefore help us gain insight and better the description of
  nonequilibrium dynamics.
  
  In this study, we focus on the nonequilibrium relaxation of an
  infinite particle system interacting via two body collisions.  We
  find that a hierarchy of scales underlies the relaxation. In
  particular, we devise an extremely simple system which exhibits
  multiscaling in infinite dimension, while in finite dimensions
  simple scaling behavior is restored. Furthermore, we show that this
  behavior extends to a broader class of collision processes.
  
  We are interested in modeling collision processes in a
  structureless, infinite dimensional space. Therefore, we place an
  infinite number of identical particles on the nodes of a completely
  connected graph. Particles are characterized by a single 
  parameter, their velocity $v$. Two-body collisions are realized by 
  choosing two particles at random and changing their velocities 
  according to $(u_1,u_2)\to (v_1,v_2)$ with
\begin{eqnarray}
\label{rule}
\pmatrix{v_1\cr v_2\cr}=\pmatrix{\gamma &1-\gamma\cr
1-\gamma&\gamma\cr}\pmatrix{u_1\cr u_2\cr}
\end{eqnarray}
and $0\leq \gamma\leq 1$. In other words, the post-collision
velocities are given by a linear combination of the pre-collision
velocities.  Both the total momentum ($u_1+u_2=v_1+v_2$) and the total
number of particles are conserved by this process. In fact, the
collision rule (\ref{rule}) is the most general linear combination
which obeys momentum conservation and Galilean invariance, i.e.,
invariance under velocity translation $v\to v-v_0$.  

Our motivation for studying this problem is inelastic collisions in
one-dimensional granular gases \cite{pkh,my,bm}. While the two
problems involve different collision rates, they share the same
trivial final state where all velocities vanish, $P(v,t)\to \delta(v)$
when $t\to\infty$ (without loss of generality, the average velocity
was set to zero by invoking the transformation $v\to v-\langle
v\rangle$). We chose to describe this work in sightly more general
terms since closely related dynamics were used in different contexts
including voting systems \cite{melzak,ff}, asset exchange processes
\cite{sps}, combinatorial processes \cite{d}, traffic flows \cite{kg},
and force fluctuations in bead packs \cite{c}. We will show that
multiscaling characterizes fluctuations in some of these problems as well.

Velocity fluctuations may be obtained via the probability
distribution function $P(v,t)$ which evolves according to the
following master equation
\begin{eqnarray}
\label{BE} 
{\partial P(v,t)\over\partial t}&=&\int_{-\infty}^\infty 
\int_{-\infty}^\infty du_1\, du_2\, P(u_1,t)P(u_2,t)\\\nonumber
&\times &\left[\delta(v-\gamma u_1-(1-\gamma)u_2)
      -\delta(v-u_2)\right].
\end{eqnarray}
The $\delta-$functions on the right-hand side reflect the collision
rule (\ref{rule}) and guarantee conservation of the number of
particles, $\int dv P(v,t)=1$, and the total momentum $\int dv\, v
P(v,t)=0$.  Eq.~(\ref{BE}) can be simplified by eliminating one of the
integrations
\begin{equation}
\label{BE1}
{\partial P(v,t)\over\partial t}+P(v,t)={1\over 1-\gamma}
\int_{-\infty}^\infty du P(u,t)P\left({v-\gamma u\over 1-\gamma},t\right).
\end{equation}
Further simplification may be achieved via the Fourier transform 
$\hat P(k,t)=\int dv\, e^{ikv}\,P(v,t)$ which obeys
\begin{equation}
\label{BEF}
{\partial \over\partial t}\,\hat P(k,t)+\hat P(k,t)=
\hat P[\gamma k,t]\,\hat P[(1-\gamma)k,t].
\end{equation}
Although the integration  is eliminated, this compact equation is
still challenging as the nonlinear term becomes nonlocal. 

Velocity fluctuations can be quantified using the moments of the velocity
distribution, \hbox{$M_n(t)=\int dv\, v^n P(v,t)$}.  The moments obey a
closed and recursive set of the ordinary differential equations. The
corresponding equations can be derived by inserting the expansion
$\hat P(k,t)=\sum_n {(ik)^n\over n!}  M_n(t)$ into Eq.~(\ref{BEF}) or
directly from Eq.~(\ref{BE}). The first few moments evolve according to 
$\dot M_0=\dot M_1=0$,  and 
\begin{eqnarray}
\dot M_2&=&-a_2M_2,\nonumber\\
\dot M_3&=&-a_3M_3,\\
\dot M_4&=&-a_4M_4+a_{24}M_2^2,\nonumber
\end{eqnarray}
with the coefficients 
\begin{equation}
\label{an}
a_n\equiv a_n(\gamma)=1-(1-\gamma)^n-\gamma^n, 
\end{equation}
and $a_{24}=6\gamma^2(1-\gamma)^2$. Integrating these 
rate equations yields $M_0=1$, $M_1=0$ and 
\begin{eqnarray}
M_2(t)&=&M_2(0)e^{-a_2t}\nonumber\\
M_3(t)&=&M_3(0)e^{-a_3t}\\
M_4(t)&=&\left[M_4(0)+3M^2_2(0)\right]e^{-a_4t}-3M_2^2(t).\nonumber
\end{eqnarray}
The asymptotic behavior of the first few moments suggests that
knowledge of the RMS fluctuation $v^*\equiv M_2^{1/2}$ is not
sufficient to characterize higher order moments since $M_3^{1/3}/v^*,
M_4^{1/4}/v^*\to\infty$, as $t\to\infty$.

This observation extends to higher order moments as
well. In general, the moments evolve according to 
\begin{equation}
\dot M_n+a_nM_n
=\sum_{m=2}^{n-2}{n\choose m}\gamma^{m}(1-\gamma)^{n-m}M_{m}M_{n-m}.
\end{equation}
Note that for \hbox{$0<\gamma<1$}, the coefficients $a_n$ satisfy
\hbox{$a_{n}<a_{m}+a_{n-m}$} when \hbox{$1<m<n-1$}.  This inequality
can be shown by introducing
\hbox{$G(\gamma)=a_{m}(\gamma)+a_{n-m}(\gamma)-a_{n}(\gamma)$} which
satisfies $G(0)=0$ and $G(\gamma)=G(1-\gamma)$.  Therefore, one needs
to show that \hbox{$G'(\gamma)=m[b_m-b_n]+(n-m)[b_{n-m}-b_n]>0$} for
\hbox{$0<\gamma< 1/2$} with \hbox{$b_n\equiv
  b_n(\gamma)=(1-\gamma)^{n-1}-\gamma^{n-1}$}. One can verify that the
$b_n$'s decrease monotonically with increasing $n$, $b_n\geq b_{n+1}$
for $n\geq 2$, therefore proving the desired inequality.  Since
moments decay exponentially, this inequality shows that the right hand
side in the above equation is negligible asymptotically. Thus, the
leading asymptotic behavior for all $n>0$ is $M_n\sim \exp(-a_n t)$.
Since the $a_n$'s increase monotonically, $a_n<a_{n+1}$, the moments
decrease monotonically in the long time limit, $M_n>M_{n+1}$.
Furthermore, in terms of the second moment one has
\begin{equation}
\label{multi}
M_{n}\propto M_2^{\alpha_n}, \qquad 
\alpha_n={1-(1-\gamma)^n-\gamma^n
\over 1-(1-\gamma)^2-\gamma^2}.
\end{equation}
While the prefactors depend on the details of the initial
distribution, the scaling exponents are universal. Therefore, the
velocity distribution does not follow a naive scaling form
$P(v,t)={1\over v^*} P({v\over v^*})$. Such a distribution would imply
the linear exponents $\alpha_n=\alpha^*_n=n/2$.  Instead, the actual
behavior is given by Eq.~(\ref{multi}) with the exponents $\alpha_n$
reflecting a multiscaling asymptotic behavior with a nontrivial
(non-linear) dependence on the index $n$.  For instance, the high
order exponents saturate, $\alpha_n\to a_2^{-1}$ for $n\to\infty$,
instead of diverging.  One may quantify the deviation from ordinary
scaling via a properly normalized set of indices
$\beta_n=\alpha_n/\alpha_n^*$ defined from $M_n^{1/n}\sim
(v^*)^{\beta_n}$. By evaluating the $\gamma=1/2$ case where
multiscaling is most pronounced, a bound can be obtained for these
indices: $7/8,31/48\leq \beta_n\leq 1$ for $n=4,6$ respectively.
Furthermore, $\beta_n\to 1-{2n-3\over 2}\gamma$ when $\gamma\to 0$
indicating that the deviation from ordinary scaling vanishes for weakly
inelastic collisions. Thus, the multiscaling behavior can be quite
subtle \cite{bk1}.

The above shows that a hierarchy of scales underlies fluctuations in 
the velocity. In parallel, a hierarchy of diverging time scales
characterizes  velocity fluctuations
\begin{equation}
M_n^{1/n}\sim \exp(-t/\tau_n), \qquad \tau_n={n\over a_n}.
\end{equation}
These time scales diverge for large $n$ according to $\tau_n\simeq n$.
Large moments reflect the large velocity tail of a distribution.
Indeed, the distribution of extremely large velocities is dominated by
persistent particles which experienced no collisions up time $t$. The
probability for such events decays exponentially with time \hbox{$P(v,t)\sim
P(v,0)\exp(-t)$} for $v\gg 1$ (alternatively, this behavior emerges
from Eq.~(\ref{BE1}) since the gain term is negligible for the tail
and hence $\dot P+P=0$).  This decay is consistent with the large
order moment decay $M_n\sim \exp(-t)$ when $n\to\infty$.

\begin{figure}
\narrowtext  
\centerline{\epsfxsize=7.5cm\epsfbox{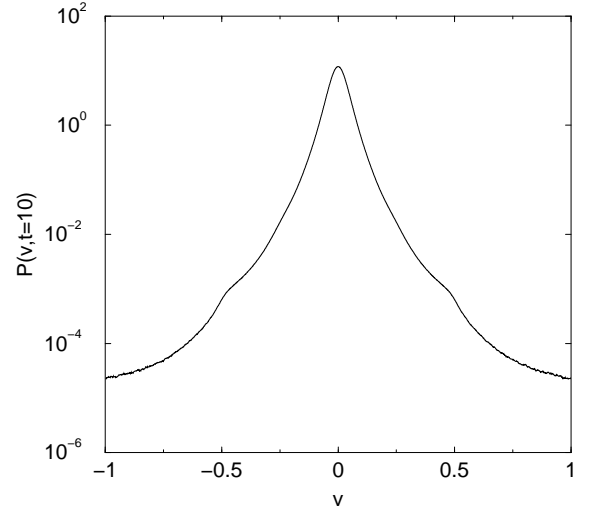}}
\caption{Development of a singularity for a compact initial distribution. 
  Shown is the probability distribution obtained by simulating the
  collision process of Eq.~(\ref{rule}) with $\gamma=1/2$. The data
  represents an average over 200 independent realization in a system
  with $10^7$ particles, starting from a uniform distribution in the
  range $[-1:1]$.}
\end{figure}

Although the leading asymptotic behavior of the moments was
established, understanding the entire distribution $P(v,t)$ remains a
challenge.  Simulations of the $\gamma=1/2$ process reveal an
interesting structure for compact distributions. Starting from a
uniform velocity distribution, $P_0(v)=1/2$ for $-1<v<1$, the
distribution loses analyticity at $v=\pm 1/2$.  Our analysis of
Eq.~(\ref{BEF}) shows that such a singularity should indeed develop at
$v=\pm 1/2$ and it additionally implies the appearance of
(progressively weaker and weaker) singularities at $v=\pm 1/4$, etc.
More generally, for an arbitrary {\em compact} initial distribution
and an arbitrary $\gamma$, the distribution $P(v,t)$ loses analyticity
for $t>0$ and develops an infinite (countable) set of singularities
whose locations depend on the arithmetic nature of $\gamma$ (e.g., it
is very different for rational and irrational $\gamma$'s). On the
other hand, unbounded distributions do not develop such singularities,
and therefore, the loss of analyticity is not necessarily 
responsible for the multiscaling behavior.

Asymptotically, our system reaches a trivial steady state 
$P(v,t=\infty)=\delta(v)$.  To examine the relation between dynamics and
statics, a non-trivial steady state can be generated by considering the 
driven version of our model \cite{wm,sbcm}. External forcing
balances dissipation due to collisions and therefore results in a
nontrivial nonequilibrium steady state.  Specifically, we assume that
in addition to changes due to collisions, velocities may also change due
to an external forcing: \hbox{${dv_j\over dt}|_{\rm heat}=\xi_j$}. We assume
standard uncorrelated white noise \hbox{$\langle\xi_i(t)
\xi_j(t')\rangle=2D\delta_{ij}\delta(t-t')$} with a zero average
$\langle \xi_j\rangle=0$.  The left hand side of the master equation
(\ref{BE1}) should therefore be modified by the diffusion term
\begin{eqnarray}
\label{BEFP} 
{\partial P(v,t)\over\partial t}\to 
{\partial P(v,t)\over\partial t}-D{\partial^2 P(v,t)\over\partial v^2}.
\end{eqnarray}
Of course, the addition of the diffusive term does not alter 
conservation of the total particle number and the total momentum, and 
one can safely work in a reference frame moving with the center of
mass velocity.

We restrict our attention to the steady state, obtained by setting the
time derivative to zero. The corresponding Fourier transform
$\hat P_\infty(k)\equiv\hat P(k,t=\infty)$ satisfies
\begin{equation}
\label{FP}
(1+Dk^2)\hat P_\infty(k)=
\hat P_\infty[\gamma k]\,\hat P_\infty[(1-\gamma)k].
\end{equation}
The solution to this functional equation which obeys the conservation
laws $\hat P_\infty(0)=1$ and $\langle v\rangle=\hat P_\infty'(0)=0$ is
found recursively 
\begin{equation}
\label{FPsol}
\hat P_\infty(k)=\prod_{i=0}^\infty\prod_{j=0}^i 
\left[1+\gamma^{2j}(1-\gamma)^{2(i-j)}Dk^2\right]^{-{i\choose j}}.
\end{equation}
To simplify this double product we take the logarithm
and transform it as follows
\begin{eqnarray}
\ln \hat P_\infty(k)
&=&-\sum_{i=0}^\infty\sum_{j=0}^j {i\choose j}\,
\ln \left[1+\gamma^{2j}(1-\gamma)^{2(i-j)}Dk^2\right]\nonumber\\ 
&=&\sum_{i=0}^\infty\sum_{j=0}^i {i\choose j}\sum_{n=1}^\infty
{(-Dk^2)^n\gamma^{2jn}(1-\gamma)^{2(i-j)n}\over n}\nonumber\\
&=&\sum_{n=1}^\infty {(-Dk^2)^n\over n}
\sum_{i=0}^\infty\sum_{j=0}^i {i\choose j}\,
\gamma^{2nj}(1-\gamma)^{2n(i-j)}\nonumber\\
&=&\sum_{n=1}^\infty {(-Dk^2)^n\over n}\sum_{i=0}^\infty
\left[\gamma^{2n}+(1-\gamma)^{2n}\right]^i.
\end{eqnarray}
The second identity follows from the series expansion $\ln
(1+q)=-\sum_{n\geq 1}n^{-1}(-q)^n$, and the forth from the binomial
identity $\sum_{j=0}^i {i\choose j}p^jq^{i-j}=(p+q)^i$. Finally, using the
geometric series $(1-x)^{-1}=\sum_{n\geq 0} x^n$, the Fourier transform
at the steady state is found
\begin{equation}
\label{pinf} 
\hat P_\infty(k)=\exp\left\{\sum_{n=1}^\infty 
{(-Dk^2)^n\over n a_{2n}(\gamma)}\right\},
\end{equation}
with $a_n(\gamma)$ given by Eq.~(\ref{an}).  The $n$th cumulant of
the steady state distribution $\kappa_n$ can be readily found from $\ln
\hat P_\infty(k)=\sum_m {(ik)^m\over m!}\kappa_m$.  Therefore, the odd
cumulants vanish while the even cumulants are simply proportional to the time
scales characterizing the exponential relaxation of the corresponding 
moments:
\begin{equation}
\kappa_{2n}={(2n-1)!\over n}D^n\tau_{2n}.
\end{equation}
Of course, the moments can be constructed from these
cumulants.  Interestingly, a direct correspondence between the steady
state characteristics and the nonequilibrium relaxation time scales is
established via the cumulants of the probability distribution.

None of the (even) cumulants vanish, thereby reflecting significant 
deviations from a Gaussian distribution. Nevertheless, for
sufficiently large velocities, one may concentrate on the small wave
number behavior. Using the inverse Fourier transform of (\ref{pinf})
one finds the tail of the distribution
\begin{equation}
P_\infty(v)\simeq \sqrt{a_2\over 4\pi D}\,
\exp\left\{-{a_2v^2\over 4D}\right\}, \qquad v\gg \sqrt{D/a_2}.  
\end{equation}
This in particular implies the large moment behavior $M_{2n}\to
(2n-1)!!(4D/a_2)^{n}$ as $n\to\infty$. 

To examine how general is the above behavior, we briefly discuss a few
generalizations and extensions of the basic model.  Relaxing Galilean
invariance, the most general momentum conserving collision rule is
\begin{eqnarray}
\label{rule1}
\pmatrix{v_1\cr v_2\cr}=\pmatrix{\gamma_1 &1-\gamma_2\cr
1-\gamma_1&\gamma_2\cr}\pmatrix{u_1\cr u_2\cr}. 
\end{eqnarray}
Following the same steps that led to (\ref{multi}) shows that when
$\gamma_1,\gamma_2\neq 0,1$ and when $M_1=0$ this process also exhibits
multiscaling with the exponents $\alpha_n=a_n/a_2$, where
$a_n(\gamma_1,\gamma_2)={1\over 2}[a_n(\gamma_1)+a_n(\gamma_2)]$.  When
$\gamma_1=1-\gamma_2=\gamma$ one recovers the model introduced by Melzak
\cite{melzak}, and when $\gamma_1=\gamma_2=\gamma$ one recovers
inelastic collisions.  Since $a_n(\gamma)=a_n(1-\gamma)$ both models
have identical multiscaling exponents. Furthermore, a multiscaling
behavior with the very same exponents $\alpha_n(\gamma)$ is also found
for the following process $(u_1,u_2)\to (u_1-\gamma u_1,v_1+\gamma
u_1)$ investigated in the context of asset distributions \cite{sps} and
headway distributions in traffic flows \cite{kg}.

One can also consider stochastic rather than deterministic collision
processes by assuming that the collision (\ref{rule1}) occurs with
probability density $\sigma_1(\gamma_1,\gamma_2)$. Our findings extend
to this model as well and the multiscaling exponents are given by the
same general expression $\alpha_n=a_n/a_2$ with \hbox{$a_n=\int
  d\gamma_1 \int d\gamma_2 \sigma(\gamma_1,\gamma_2)
  a_n(\gamma_1,\gamma_2 )$}. In particular, for completely random
inelastic collisions, i.e., $\sigma\equiv 1$ and
$\gamma_1=\gamma_2=\gamma$, one finds $a_n={n-1\over n+1}$ and hence
$\alpha_n=3{n-1\over n+1}$.

So far, we discussed only two-body interactions. We therefore consider
$N$-body interactions where a collision is symbolized by
$(u_1,\ldots,u_N)\to (v_1,\ldots,v_N)$. We consider a generalization
of the $\gamma={1\over 2}$ two-body case where the post-collision
velocities are all equal.  Momentum conservation implies $v_i=\bar
u=N^{-1}\sum u_i$. The master equation is a straightforward
generalization of the two-body case and we merely quote the moment
equations
\begin{equation}
\dot M_n+a_nM_n=
N^{-n}\sum_{n_i\neq 1} {n\choose {n_1\ldots n_N}} M_{n_1}\cdots M_{n_N}
\end{equation}
with $a_n=1-N^{1-n}$. Using the inequality \hbox{$a_n<a_m+a_{n-m}$}
for all \hbox{$1<m<n-1$}, and its kin like
\hbox{$a_n<a_{m_1}+a_{m_2}+a_{n-m_1-m_2}$} for all
\hbox{$1<m_1,m_2<n-1$}, etc., we find that the right-hand side of the
above equation remains asymptotically negligible.  Therefore, \hbox{$M_n\sim
e^{-a_n t}$} and
\begin{equation}
M_n\sim M_2^{\alpha_n},\qquad
\alpha_n={1-N^{1-n}\over 1-N^{-1}}.
\end{equation}
Thus, this $N$-body ``averaging'' process exhibits multiscaling
asymptotic behavior as well.

Thus far, we considered the behavior on a mean field level, i.e., in
an infinite dimensional space. It is natural to consider the
finite-dimensional counterpart. Specifically, we assume that particles
reside on a $d$ dimensional lattice and that only nearest neighbors
interact. Here, the above dynamics is essentially equivalent to a
diffusion process \cite{bk}.  As a result, the underlying correlation
length is diffusive, $L(t)\sim t^{1/2}$. Within this correlation
length the velocities are ``well mixed'' and momentum conservation
therefore implies that $v\sim L^{-d/2}\sim t^{-d/4}$.  Indeed, the
infinite dimension limit is consistent with the above exponential
decay.  Furthermore, an exact solution for moments of arbitrary order
is possible \cite{bk}.  We do not detail it here and simply quote 
that ordinary scaling is restored $M_n\sim t^{-n/4}$, i.e.
$\alpha_n=\alpha_n^*=n/2$. Thus, spatial correlations
counter the mechanism responsible for multiscaling. 

In summary, we have investigated inelastic collision processes in
infinite dimension.  We have shown that such systems are characterized
by multiscaling, or equivalently by an infinite hierarchy of diverging
time scales.  Multiscaling holds for several generalizations of the
basic model including stochastic collision models and even processes
which do not obey Galilean invariance. In this latter case, however,
multiscaling is restricted to situations with zero total momentum.
This perhaps explains why multiscaling asymptotic behavior was
overlooked in previous studies \cite{melzak,sps}. Another explanation
is that this behavior may be difficult to detect from numerical
simulations. Indeed, in other problems such as multidimensional
fragmentation \cite{bk1}, and in fluid turbulence, low order moments
deviate only slightly from the normal scaling expectation.

There are a number of extensions of this work which are worth
pursuing. We have started with a simplified model of a 1D granular gas
with a velocity independent collision rate. One possibility is to
approximate the collision rate with the RMS velocity fluctuation.
This leads to the algebraic decay $M_n\sim t^{-2\alpha_n}$ with
$\alpha_n$ given by Eq.~(\ref{multi}) and in particular, Haff's
cooling law $T=M_2\sim t^{-2}$ is recovered \cite{pkh}. Our numerical
studies indicate that when velocity dependent collision rates are
implemented, ordinary scaling behavior is restored.  One may also use
this model as an approximation for inelastic collisions in higher
dimensions as well, following the Maxwell approximation in
kinetic theory \cite{ernst,bk2}.

\medskip 
This research was supported by the DOE (W-7405-ENG-36), NSF
(DMR9632059), and ARO (DAAH04-96-1-0114).

\end{multicols}

\begin{thebibliography}{99}

\bibitem{pkh} 
      P.~K. Haff, J. Fluid Mech. {\bf 134}, 401 (1983).
\bibitem{my} 
      S.~McNamara and W.~R. Young, Phys Fluids A {\bf 4}, 496 (1992).
\bibitem{bm} 
      B.~Bernu and R.~Mazighi, J. Phys. A {\bf 23}, 5745 (1990). 
\bibitem{melzak} 
      Z.~A.~Melzak, {\it Mathematical Ideas, Modeling and Applications,
      Volume II of Companion to Concrete Mathematics} (Wiley, New York,
      1976), p.\ 279.  
\bibitem{ff} 
      P.~A.~Ferrari and L.~R.~G.~Fontes, El. J. Prob. {\bf 3}, Paper
      no.~6 (1998).
\bibitem{sps} 
      S.~Ispolatov, P.~L.~Krapivsky, and S.~Redner, 
      Eur. Phys. J. B {\bf 2}, 267 (1998).
\bibitem{d} 
      D.~Aldous and P.~Diaconis, Prob. Theory Relat. Fields {\bf 103}, 
      199 (1995).
\bibitem{kg} 
      J.~Krug and J.~Garisa, {\it cond-mat/9909034}.
\bibitem{c} 
      S.~N.~Coppersmith, C.-h.~Liu, S.~Majumdaar, O.~Narayan, and
      T.~Witten, Phys. Rev. E {\bf 53}, 4673 (1996).
\bibitem{bk1} 
      P.~L.~Krapivsky and E.~Ben-Naim, Phys. Rev. E {\bf 50}, 3502
      (1994); E.~Ben-Naim and P.~L.~Krapivsky, Phys. Rev. Lett. 
      {\bf 76}, 3234 (1996).
\bibitem{wm} D.~R.~M.~Williams and
      F.~C.~MacKintosh, Phys. Rev. E {\bf 54}, 9 (1996).
\bibitem{sbcm}
      M.~R.~Swift, M.~Boamf\v a, S.~J.~Cornell, and A.~Maritan, Phys. Rev.
      Lett.  {\bf 80}, 4410 (1998).
\bibitem{bk}
      E.~Ben-Naim and P.~L.~Krapivsky, in preparation.
\bibitem{ernst}
      M.~H.~Ernst, Phys. Rep. {\bf 78}, 1 (1981). 
\bibitem{bk2}
      E.~Ben-Naim and P.~L.~Krapivsky, 
      Phys. Rev. E {\bf 59}, 7000 (1999). 

\end{thebibliography}
\end{document}